# ЦИФРОВАЯ ПЕРЕДИСКРЕТИЗАЦИЯ С ПОМОЩЬЮ СТРУКТУР ФАРРОУ

**Бахолдин Н.В., Чернienko В.А., Бахурин С.А., к.т.н**

*МФТИ*
*141701, Россия, г. Долгопрудный, Институтский переулок, 9*
*bakholdin.nv@phystech.edu*

**Аннотация.** Фильтры Фарроу позволяют разрабатывать универсальные передискретизаторы для коррекции дробной задержки и преобразования частоты дискретизации. Использование кубической интерполяции Лагранжа в фильтре Фарроу не позволяет обеспечить полосу обработки более 0,4 частоты дискретизации Fд, а также не гарантирует гладкость импульсной характеристики интерполятора. В данной работе предлагаются структуры фильтра Фарроу, основанные на сплайнах Эрмита 3-го, 5-го и 7-го порядка с использованием дифференциатора высокого порядка. Предложенные фильтры увеличивают полосу обработки до 0,8Fд за счет повышения порядка полинома с дополнительными ограничениями на непрерывность производных в узлах интерполяции.

**Ключевые слова:** *Цифровая обработка сигналов; фильтр Фарроу; коррекция дробной задержки; цифровая передискретизация.*

## Введение

Развитие систем связи привело к появлению большого числа стандартов. Необходимость увеличения скорости передачи информации требует повышения порядков модуляции, что, в свою очередь, накладывает особые требования на точность синхронизации. Использование независимых задающих генераторов в передатчике и приемнике системы связи приводит к различиям в частотах дискретизации сигналов. Кроме того, часто возникает задача компенсации дробной задержки, в пределах интервала дискретизации [1, 2]. Фильтр, способный компенсировать произвольную задержку, может быть использован для синхронизации задающих генераторов, а также для преобразования частоты дискретизации в дробное число раз, P/Q. Если $P > 1$ и $Q = 1$, то в этом случае получим цифровой интерполятор, который увеличивает частоту дискретизации в P раз.

Стоит отметить, что когда мы устремляем $P \to \infty$, мы переходим от цифровой области к аналоговой, то есть $h_n(t) \to h(t)$.

Одна из универсальных архитектур цифровых передискретизаторов — это фильтры Фарроу, основанные на схеме интерполяции Лагранжа [3]. Несмотря на свою универсальность, Лагранжева интерполяция имеет ряд недостатков и не может быть использована в высококачественных передискретизаторах в широкой полосе частот.

В данной работе предложены структуры фильтров Фарроу, основанные на сплайнах Эрмита [4] с использованием широкополосных дифференцирующих фильтров, что приводит к заметному улучшению качества передискретизации по сравнению с конфигурациями, основанными на интерполяции полиномами Лагранжа.

## Фильтр Фарроу на основе сплайнов Эрмита

Мы можем увеличить порядок полинома Эрмита, накладывая дополнительные ограничения на гладкость импульсной характеристики фильтра Фарроу.

Оценка производных может быть выполнена с использованием дополнительного фиксированного дифференцирующего КИХ-фильтра (рис. 2), который может быть оптимизирован для аппаратной реализации [7, 8].

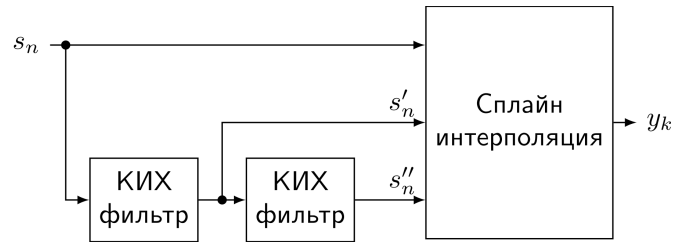

Рисунок 1. Фильтр Фарроу с дифференцирующими фильтрами

**Фильтр Фарроу на основе кубических сплайнов Эрмита**

Дополнительные ограничения на производные сплайна Эрмита позволяют увеличить его порядок (рис. 3), что улучшает свойства фильтра Фарроу.

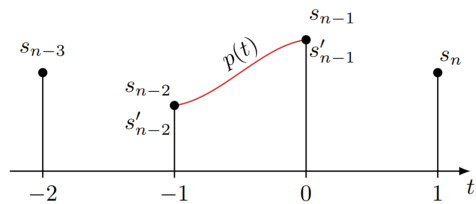

Рисунок 2. Расчет кубического сплайна Эрмита

Полином рассчитывается в интервале $t = [-1 \ldots 0]$. Ниже приведены уравнения для вычисления кубического сплайна Эрмита:

$$\begin{cases} p(0) = s_{n-1}, & p(-1) = s_{n-2}, \\ p'(0) = s'_{n-1}, & p'(-1) = s'_{n-2}; \end{cases} \quad (1)$$

где $p'(0)$ и $p'(-1)$ — оценки первых производных в точках $t = 0$ и $t = -1$.

Коэффициенты полинома будут вычисляться путём решения системы линейных уравнений:

$$\begin{cases} a_0 + a_1 \cdot 0 + a_2 \cdot 0^2 + a_3 \cdot 0^3 = s_{n-1}, \\ a_0 + a_1 \cdot (-1) + a_2 \cdot (-1)^2 + a_3 \cdot (-1)^3 = s_{n-2}, \\ a_0 \cdot 0 + a_1 + 2a_2 \cdot 0 + 3a_3 \cdot 0^2 = s'_{n-1}, \\ a_0 \cdot 0 + a_1 + 2a_2 \cdot (-1) + 3a_3 \cdot (-1)^2 = s'_{n-2}. \end{cases} \quad (2)$$

В матричной форме уравнение (2) имеет вид:

$$\mathbf{Ma} = \mathbf{s} = \begin{bmatrix} 1 & 0 & 0 & 0 \\ 1 & -1 & 1 & -1 \\ 0 & 1 & 0 & 0 \\ 0 & 1 & -2 & 3 \end{bmatrix} \cdot \begin{bmatrix} a_0 \\ a_1 \\ a_2 \\ a_3 \end{bmatrix} = \begin{bmatrix} s_{n-1} \\ s_{n-2} \\ s'_{n-1} \\ s'_{n-2} \end{bmatrix}. \tag{3}$$

Коэффициенты **a** могут быть найдены:

$$\mathbf{a} = \begin{bmatrix} 1 & 0 & 0 & 0 \\ 0 & 0 & 1 & 0 \\ -3 & 3 & 2 & 1 \\ -2 & 2 & 1 & 1 \end{bmatrix} \cdot \begin{bmatrix} s_{n-1} \\ s_{n-2} \\ s'_{n-1} \\ s'_{n-2} \end{bmatrix}. \tag{4}$$

После упрощения коэффициенты кубического сплайна Эрмита имеют вид:

$$\begin{cases} a_0 = s_{n-1}, & a_1 = s'_{n-1}, \\ a_2 = 3(s_{n-2} - s_{n-1}) + 2s'_{n-1} + s'_{n-2}, \\ a_3 = 2(s_{n-2} - s_{n-1}) + s'_{n-1} + s'_{n-2}. \end{cases} \tag{5}$$

Как видно, фильтр Фарроу на основе кубического сплайна Эрмита требует умножения на 2 и 3, которые могут быть реализованы без использования умножителей.

Оценка производных для вычисления коэффициентов сплайна Эрмита обеспечивает непрерывность производной импульсной характеристики фильтра-интерполятора при $P \to \infty$, как будет показано в следующих разделах.

**Фильтр Фарроу на основе сплайнов Эрмита 5 порядка**

Процесс построения сплайна Эрмита 5 порядка изображен на рисунке 4а.

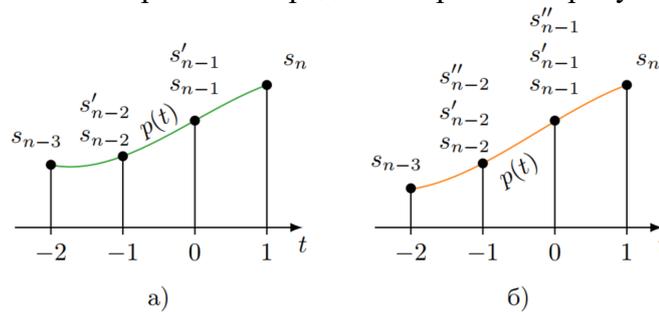

Рисунок 3. а) Построение сплайна Эрмита 5 порядка, б) Построение сплайна Эрмита 7 порядка

Полином 5 порядка проходит через 4 точки с дополнительными ограничениями на гладкость производных в узлах интерполяции:

$$\begin{cases} p(1) = s_n, & p(0) = s_{n-1}, \\ p(-1) = s_{n-2}, & p(-2) = s_{n-3}, \\ p'(0) = s'_{n-1}, & p'(-1) = s'_{n-2}. \end{cases} \qquad (6)$$

Для потроения полинома необходимо рассчитать 6 коэффициентов, используя приведенное выше уравнение, записанное в матричнй форме:

$$\begin{bmatrix} 1 & 1 & 1 & 1 & 1 & 1 \\ 1 & 0 & 0 & 0 & 0 & 0 \\ 1 & -1 & 1 & -1 & 1 & -1 \\ 1 & -2 & 4 & -8 & 16 & -32 \\ 0 & 1 & 0 & 0 & 0 & 0 \\ 0 & 1 & -2 & 3 & -4 & 5 \end{bmatrix} \cdot \begin{bmatrix} a_0 \\ a_1 \\ a_2 \\ a_3 \\ a_4 \\ a_5 \end{bmatrix} = \begin{bmatrix} s_n \\ s_{n-1} \\ s_{n-2} \\ s_{n-3} \\ s'_{n-1} \\ s'_{n-2} \end{bmatrix}. \qquad (7)$$

Коэффициенты $a$ могут быть найдены как:

$$\begin{bmatrix} 0 & 1 & 0 & 0 & 0 & 0 \\ 0 & 0 & 0 & 0 & 1 & 0 \\ \frac{1}{6} & -\frac{11}{4} & \frac{5}{2} & \frac{1}{12} & \frac{3}{2} & 1 \\ \frac{5}{12} & -\frac{3}{4} & \frac{1}{4} & \frac{1}{12} & -\frac{1}{2} & \frac{1}{2} \\ \frac{1}{3} & \frac{7}{4} & -2 & -\frac{1}{12} & -\frac{3}{2} & -1 \\ \frac{1}{12} & \frac{3}{4} & -\frac{3}{4} & -\frac{1}{12} & -\frac{1}{2} & -\frac{1}{2} \end{bmatrix} \cdot \begin{bmatrix} s_n \\ s_{n-1} \\ s_{n-2} \\ s_{n-3} \\ s'_{n-1} \\ s'_{n-2} \end{bmatrix}. \qquad (8)$$

Обратная матрица также содержит повторяющиеся элементы. После упрощения количество умножителей матрично-векторного произведения получаем:

$$\begin{cases} a_0 = s_{n-1}, \\ a_1 = s'_{n-1}, \\ a_2 = \frac{1}{12}(2s_n + s_{n-3}) + \frac{1}{2}(5s_{n-2} + 3a_1) - \frac{11}{4}a_0 + s'_{n-2}, \\ a_3 = \frac{1}{12}(5s_n + s_{n-3}) - \frac{1}{4}(3a_0 - s_{n-2}) - \frac{1}{2}(a_1 - s'_{n-2}), \\ a_4 = \frac{1}{2}(s_n + s_{n-2}) - a_0 - a_2, \\ a_5 = \frac{1}{2}(s_n - s_{n-2}) - a_1 - a_3, \end{cases} \qquad (9)$$

Так, для аппаратной реализации необходимо использовать 7 умножителей: $\frac{1}{12}, \frac{1}{4}, \frac{1}{2}$, 2, 3, 5, 11, из которых три могут быть реализованы как регистры сдвига в целочисленной арифметике.

**Фильтр Фарроу на основе сплайнов Эрмита 7 порядка**

Для вычисления сплайна седьмого порядка необходимо записать восемь уравнений: четыре из них задают значения полинома, остальные — ограничения на первые и вторые производные, как это показано на рисунке 4б.

Система линейных уравнений имеет вид:

$$\begin{cases} p(1) = s_n, & p(0) = s_{n-1}, \\ p(-1) = s_{n-2}, & p(-2) = s_{n-3}, \\ p'(0) = s'_{n-1}, & p'(-1) = s'_{n-2}, \\ p''(0) = s''_{n-1}, & p''(-1) = s''_{n-2}; \end{cases} \tag{10}$$

Система линейных уравнений (10) может быть представлена в матричной форме $\mathbf{Ma} = \mathbf{s}$:

$$\mathbf{M} = \begin{bmatrix} 1 & 0 & 0 & 0 & 0 & 0 & 0 & 0 \\ 1 & -1 & 1 & -1 & 1 & -1 & 1 & -1 \\ 1 & 1 & 1 & 1 & 1 & 1 & 1 & 1 \\ 1 & -2 & 4 & -8 & 16 & -32 & 64 & -128 \\ 0 & 1 & 0 & 0 & 0 & 0 & 0 & 0 \\ 0 & 1 & -2 & 3 & -4 & 5 & -6 & 7 \\ 0 & 0 & 2 & 0 & 0 & 0 & 0 & 0 \\ 0 & 0 & 2 & -6 & 12 & -20 & 30 & -42 \end{bmatrix},$$

$$\mathbf{a} = \begin{bmatrix} a_0 & a_1 & a_2 & a_3 & a_4 & a_5 & a_6 & a_7 \end{bmatrix}^T,$$

$$\mathbf{s} = \begin{bmatrix} s_n & s_{n-1} & s_{n-2} & s_{n-3} & s'_{n-1} & s'_{n-2} & s''_{n-1} & s''_{n-2} \end{bmatrix}^T, \tag{11}$$

Обратная матрица $\mathbf{M^{-1}}$:

$$\mathbf{M^{-1}} = \begin{bmatrix} 1 & 0 & 0 & 0 & 0 & 0 & 0 & 0 \\ 0 & 0 & 0 & 0 & 1 & 0 & 0 & 0 \\ 0 & 0 & 0 & 0 & 0 & 0 & \frac{1}{2} & 0 \\ \frac{69}{8} & -\frac{35}{4} & \frac{1}{12} & \frac{1}{24} & -\frac{21}{4} & -\frac{7}{2} & \frac{5}{4} & -\frac{1}{2} \\ \frac{35}{4} & -\frac{69}{8} & \frac{7}{24} & \frac{1}{12} & -\frac{9}{2} & -\frac{17}{4} & \frac{1}{2} & -\frac{3}{4} \\ -6 & \frac{45}{8} & \frac{3}{8} & 0 & 3 & \frac{9}{4} & -1 & \frac{1}{4} \\ -\frac{37}{4} & \frac{73}{8} & \frac{5}{24} & -\frac{1}{12} & \frac{9}{2} & \frac{17}{4} & -1 & \frac{3}{4} \\ -\frac{21}{8} & \frac{21}{8} & \frac{1}{24} & -\frac{1}{24} & \frac{5}{4} & \frac{5}{4} & -\frac{1}{4} & \frac{1}{4} \end{bmatrix}. \tag{12}$$

Уравнения для коэффициентов a после сокращения сложности:

$$\begin{cases} a_0 = s_{n-1}, \qquad a_1 = s'_{n-1}, \qquad a_2 = \tfrac{1}{2}s''_{n-1}, \\ a_3 = \tfrac{1}{24}(207a_0 + s_{n-3} + 2s_n) - \tfrac{1}{4}(35s_{n-2} + \\ \qquad + 21a_1 + 14s'_{n-2} - 5s''_{n-1} + 2s''_{n-2}), \\ a_4 = \tfrac{1}{2}(s_n + s_{n-2}) - a_0 - a_2 - a_6, \\ a_5 = \tfrac{3}{8}(s_n + 15s_{n-2}) - 6a_0 + 3a_1 + \\ \qquad + \tfrac{1}{4}(9s'_{n-2} + s''_{n-2}) - s''_{n-1}, \\ a_6 = \tfrac{1}{24}(5s_n + 219s_{n-2} - 2s_{n-3}) + \\ \qquad + \tfrac{1}{4}(-37a_0 + 18a_1 + 17s'_{n-2} + 3s''_{n-2}) - s''_{n-1}, \\ a_7 = \tfrac{1}{24}(s_n - s_{n-3}) - \tfrac{21}{8}(a_0 - s_{n-2}) + \\ \qquad + \tfrac{1}{4}(5(a_1 + s'_{n-2}) - s''_{n-1} + s''_{n-2}). \end{cases} \qquad (13)$$

В результате требуется 18 умножителей с коэффициентами $\tfrac{1}{24}, \tfrac{1}{8}, \tfrac{1}{4}, \tfrac{1}{2}, 2, 3, 5, 6, 9, 14, 15, 17, 18, 21, 35, 37, 207, 219$, четыре из которых могут быть реализованы как регистры сдвига в целочисленной арифметике.

**Сравнение характеристик фильтра Фарроу на основе полинома Лагранжа и сплайнов Эрмита**

Характеристики фильтров Фарроу, основанные на различных сплайнах, показаны на рисунке 5.

Использование ограничений на производные обеспечивает непрерывность импульсной характеристики $h(t)$.

Уровень боковых лепестков уменьшился с −30 дБ до −36 дБ при использовании кубического сплайна Эрмита вместо полинома Лагранжа.

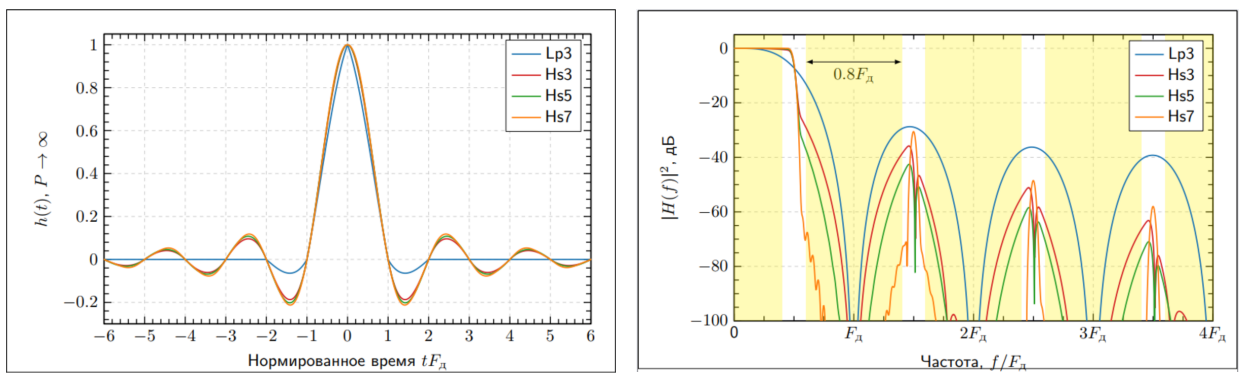

Рисунок 4. Импульсная (а) и частотная (б) характеристика фильтра Фарроу на основе кубического полинома Лагранжа (Lp3) и сплайнов Эрмита 3 порядка (Hs3), 5 порядка (Hs5) и 7 порядка (Hs7); порядок дифференцирующего фильтра равен 32

Кроме того, дальнейшее увеличение порядка сплайна Эрмита приводит к дополнительному уменьшению уровня боковых лепестков.

Частотная ось на рисунке 5б нормирована по входной частоте сигнала Fд, и после интерполяции в $P = 8$ раз частота Найквиста становится 4Fд, что и показано на графике.

На рисунке 5б желтым цветом выделена полоса обработки 0,8Fд до интерполяции в 8 раз. Мы видим, что фильтр Фарроу, основанный на сплайнах 7-го порядка, использует производные второго порядка и обеспечивает подавление копий спектра до 65 дБ в полосе обработки 0,8Fд.

**Фильтры дробной задержки на базе сплайнов Эрмита**

Универсальность ресемплеров Фарроу достигается из-за возможности интерполяции в непрерывном времени и получением значений сигнала для произвольной дробной задержки. Поэтому качество интерполяции при применении дробной задержки влияет на качество передискретизации.

Групповая задержка фильтров Фарроу, основанных на полиноме Лагранжа и кубических сплайнах Эрмита, показана на рисунке 9 для различных значений дробных задержек от 0 до 1. Значение $D = N_{FIR}/2$ было вычтено из групповой задержки $\tau(f)$, для отображения в интервале [0, 1].

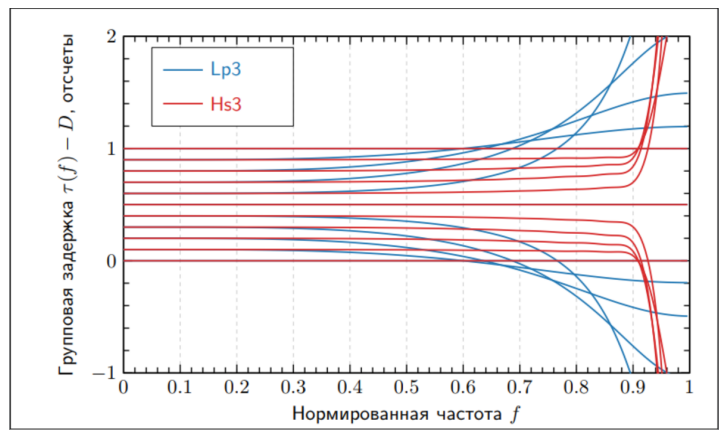

Рисунок 6. Групповая задержка фильтра Фарроу с компенсацией дробной задержки на основе кубического полинома Лагранжа (Lp3) и кубического сплайна Эрмита (Hs3), дифференцирующий фильтр 48 порядка

Показано, что фильтры Фарроу, основанные на кубическом сплайне Эрмита, имеют полосу постоянной групповой задержки 0,8Fд, в то время как фильтр, основанный на полиноме Лагранжа, имеет полосу постоянной групповой задержки не более 0,4Fд.

Расширение полосы достигается за счет применения производных, полученных на выходе широкополосного дифференцирующего КИХ-фильтра 48 порядка.

**Заключение**

В данной статье рассмотрены различные методы проектирования фильтра Фарроу с использованием полинома Лагранжа и сплайнов Эрмита. Недостатки фильтра Фарроу с полиномом Лагранжа, такие как негладкая импульсная характеристика (что приводит к плохим фильтрующим свойствам), могут быть улучшены с использованием сплайнов Эрмита с вычислением первых и вторых производных с помощью дополнительных дифференцирующих КИХ-фильтров высокого порядка. Также КИХ фильтры высокого порядка позволяют значительно расширить полосу обработки сигнала.

Уровень боковых лепестков уменьшился с −30 дБ до −36 дБ при использовании кубического сплайна Эрмита вместо полинома Лагранжа. При этом обеспечивается гладкость импульсной характеристики (рис. 5а). Более того, дальнейшее увеличение порядка сплайна Эрмита приводит к дополнительному уменьшению уровня боковых лепестков.

Исследования показали, что фильтры Фарроу на базе кубического сплайна Эрмита обеспечивают полосу постоянной групповой задержки 0,8Fд, в то время как для фильтров, использующих полиномы Лагранжа, эта полоса ограничена значением 0,4Fд. Расширение полосы достигается за счет использования производных, полученных с помощью широкополосного дифференцирующего КИХ-фильтра 48 порядка.

*Литература*